\documentclass[12pt]{article}
\usepackage{latexsym}
\usepackage{amssymb}
\catcode`\@=11
\newcount\hour
\newcount\minute
\newtoks\amorpm \hour=\time\divide\hour by 60\minute
=\time{\multiply\hour by 60 \global\advance\minute by-\hour}
\edef\standardtime{{\ifnum\hour<12 \global\amorpm={am}%
        \else\global\amorpm={pm}\advance\hour by-12 \fi
        \ifnum\hour=0 \hour=12 \fi
        \number\hour:\ifnum\minute<10
        0\fi\number\minute\the\amorpm}}
\edef\militarytime{\number\hour:\ifnum\minute<10
0\fi\number\minute}
\def\draftlabel#1{{\@bsphack\if@filesw {\let\thepage\relax
   \xdef\@gtempa{\write\@auxout{\string
      \newlabel{#1}{{\@currentlabel}{\thepage}}}}}\@gtempa
   \if@nobreak \ifvmode\nobreak\fi\fi\fi\@esphack}
        \gdef\@eqnlabel{#1}}
\def\@eqnlabel{}
\def\@vacuum{}
\def\marginnote#1{}
\def\draftmarginnote#1{\marginpar{\raggedright\scriptsize\tt#1}}
\def\draft{
        \pagestyle{plain}
        \overfullrule=2pt
        \oddsidemargin -.1truein
        \def\@oddhead{\sl \phantom{\today\quad\militarytime} \hfil
        \smash{\Large\sl DRAFT} \hfil \today\quad\militarytime}
        \let\@evenhead\@oddhead
        \let\label=\draftlabel
        \let\marginnote=\draftmarginnote
        \def\ps@empty{\let\@mkboth\@gobbletwo
        \def\@oddfoot{\hfil \smash{\Large\sl DRAFT} \hfil}
        \let\@evenfoot\@oddhead}
        \def\@eqnnum{(\theequation)\rlap{\kern\marginparsep\tt\@eqnlabel}%
        \global\let\@eqnlabel\@vacuum}  }

\renewcommand{\theequation}{\thesection.\arabic{equation}}
\renewcommand{\thefootnote}{\fnsymbol{footnote}}

\def\appendix#1{\addtocounter{section}{1}\setcounter{equation}{0}
\renewcommand{\thesection}{\Alph{section}}
\section*{Appendix \thesection\protect\indent \parbox[t]{11.15cm}{#1}}
\addcontentsline{toc}{section}{Appendix \thesection\ \ \ #1}}

\def \bi{\bibitem}
\def \la {\label}

\def \b {\beta}

\def\r2{\sqrt{2}}
\def\be{\begin{equation}}
\def\ee{\end{equation}}

\def\nat {{\natural}}

\catcode`\@=11

\def\bea{\begin{eqnarray}}
\def\eea{\end{eqnarray}}
\def\beann{\begin{eqnarray*}}
\def\eeann{\end{eqnarray*}}
\def\beq{\begin{equation}}
\def\eeq{\end{equation}}
\def\ba{\begin{array}}
\def\ea{\end{array}}
\def\ben{\begin{enumerate}}
\def\een{\end{enumerate}}

 \def \la {\label}
 \def\be{\begin{equation}}
\def\ee{\end{equation}}

\def \la {\label}

\def \r {\rho}

\font\mybb=msbm10 at 11pt

\def\bb#1{\hbox{\mybb#1}}

\def\bR {\bb{R}}

\def \ee {\epsilon}

\def \g {\gamma}
\def \bi{\bibitem}
\def\a{\alpha }

\def \r {\rho}

\def \g {\gamma}

\def \b {\beta}

\def\be{\begin{equation}}
\def\ee{\end{equation}}

\def \bi {\bibitem}
\def \la{\label}

\begin{document}
\date{November 2002}
\begin{titlepage}
\begin{center}
\hfill hep-th/0601164\\

\vspace{1.5cm}
{\Large \bf  Kappa symmetry, generalized calibrations and spinorial geometry}\\[.2cm]

\vspace{1.5cm}
 {\large    G.~Papadopoulos  and P.~Sloane}

 \vspace{0.5cm}

Department of Mathematics\\
King's College London\\
Strand\\
London WC2R 2LS
\end{center}

\vskip 1.5 cm
\begin{abstract}

We extend the spinorial geometry techniques  developed for the solution
of supergravity Killing spinor equations to the  kappa
symmetry condition for supersymmetric brane probe
configurations in any supergravity background. In particular, we construct the linear systems associated with
the kappa symmetry projector of M- and type II branes acting on any Killing spinor. As an example, we  show that  static supersymmetric
 M2-brane configurations  which admit a Killing spinor
 representing the $SU(5)$ orbit of $Spin(10,1)$ are
generalized almost hermitian calibrations and the embedding map is pseudo-holomorphic.
We also present a bound for the Euclidean action of  M- and type II branes embedded
in a supersymmetric background with non-vanishing fluxes. This leads to  an extension
of the definition of generalized calibrations which allows for the presence of non-trivial
Born-Infeld type of fields in the brane actions.

\end{abstract}
\end{titlepage}
\newpage
\setcounter{page}{1}
\renewcommand{\thefootnote}{\arabic{footnote}}
\setcounter{footnote}{0}

\setcounter{section}{0}
\setcounter{subsection}{0}
\newpage

The supersymmetry condition
for a probe brane propagating in a supersymmetric supergravity background with a Killing spinor $\epsilon$ is
\bea
{ \Gamma}\epsilon=\epsilon~,
\la{kappapro}
\eea
where $\Gamma$ is the kappa symmetry projector, $\Gamma^2=1$ and ${\rm tr}\, \Gamma=0$. This condition was found for M2-branes
in \cite{pope}, extended to Euclidean M-branes and type II NS$\otimes$NS branes in \cite{strominger} and generalized
to D-branes with non-vanishing worldvolume fluxes in \cite{ericgeorge}.
The projector $\Gamma$ is an endomorphism of a Clifford algebra and depends on the embedding map $X$ of
the brane world-volume into spacetime and on the spacetime frame $e$ \cite{green}-\cite{bandos}. For
some branes, it also depends
 on other world-volume fields
and on some of the spacetime gauge potentials,
e.g. for D-branes it depends on the Born-Infeld field $F$ and on the NS$\otimes$NS two-form gauge potential $B$ and so
$\Gamma=\Gamma(e, X, F, B)$. The kappa symmetry projectors that we use in this paper are summarized in \cite{ericgeorge}.
Suppose that  a supergravity background admits $N$ Killing spinors, $\{\epsilon_i; i=1,\dots,N\}$,
${\cal D}\epsilon_i=0$, ${\cal A}\epsilon_i=0$, where ${\cal D}$ is the supercovariant connection associated with the
gravitino supersymmetry transformation and ${\cal A}$ are algebraic
conditions associated with the supersymmetry transformations of the rest of the fermions.  The solutions
of the kappa symmetry condition can be written as
$\epsilon_r=\sum_i u_{ri} \epsilon_i$, where $u$ are real constant coefficients. Supersymmetric brane
configurations are those for which (\ref{kappapro}) admits as solutions $K$, $0<K\leq N$, linearly  independent
Killing spinors. In such a case, the brane probe preserves $K$ spacetime supersymmetries.

In \cite{jug}, a spinorial geometry method  has been proposed   to solve the Killing spinor
equations of supergravity theories. This method turns the Killing spinor equations to a linear system
of algebraic equations and a parallel transport equation which however do not contain spacetime gamma matrices.
The linear system can be solved to determine some of the fluxes in terms of the geometry and to find the conditions
on the geometry of spacetime imposed by supersymmetry. The advantage of the method is that the calculation can be
done rather efficiently and  in generality, i.e. without  making
ansatze for the spacetime geometry, supergravity fluxes or Killing spinors.  The spinorial geometry is based on the
following ingredients:

\begin{itemize}

\item a description of spinors in terms of forms,

\item an oscillator basis in the space of spinors,

\item the use of a gauge symmetry of the Killing spinor equations to bring the Killing spinors into a normal form~.

\end{itemize}

The latter is a key ingredient because it can be used to solve the Killing spinor equations rather efficiently
for a small number of supersymmetries and simplifies the linear systems in all cases.

Although  the spinorial geometry method has been initially developed to solve the supergravity Killing spinor equations,
it can be easily adapted
to other spinorial problems and in particular to solve the superymmetry condition for brane probes (\ref{kappapro})
without using ansatze for the fields.
The first two ingredients of the method are based on well known properties of  spinors and they have been extensively
explained and applied in the context of ten- and eleven-dimensional supergravities \cite{jug}-\cite{ugplgpa},  so we shall not elaborate
upon these here.
It remains to find the gauge symmetry of the  supersymmetry condition (\ref{kappapro}) which can be used
in the context of spinorial geometry. These are defined as those local transformations, automorphisms of the spinor bundle, with values
in $GL(\Delta)$, $\Delta$ is a spinor representation\footnote{The spinor representation $\Delta$ depends
on the supergravity theory under investigation.},  that act on the spinor $\epsilon$ and leave the
form of the supersymmetry condition invariant, i.e.
\bea
g^{-1} \Gamma(e, X, {\cal F}) g=\Gamma(e^g, X^g, {\cal F}^g)
\eea
where ${\cal F}$ denotes collectively the remaining fields that are required for the definition of  $\Gamma$ and  $e\rightarrow e^g$, $X\rightarrow
X^g$ and ${\cal F}\rightarrow {\cal F}^g$ is an induced transformation on the fields.
It is well-known that  $\Gamma$ is invariant
under worldvolume orientation preserving diffeomorphisms. However the gauge
transformations that are relevant in the context of spinorial geometry are the local $Spin\subset GL$ transformations.
In particular for M-branes, the gauge group is $Spin(10,1)$ which is the same as the gauge group of the Killing spinor
equations of eleven-dimensional supergravity. For both M2- and M5-branes, the spin transformations
$g\in Spin(10,1)$ induce  Lorentz rotations on the spacetime frame $e$. This can be easily seen by a direct
inspection  of the kappa symmetry projectors of the M2-and M5-branes, in the form given
in \cite{eric} and in \cite{howe}, respectively, see also (\ref{kappam2}).

An inspection of the kappa symmetry projectors of type II branes  \cite{green, pope, eric, cederwall} reveals   that the gauge
symmetry of the kappa symmetry
projectors is $Spin(9,1)$. This is again compensated by a Lorentz transformation
of the spacetime frame. This gauge symmetry is the same as that of the Killing spinor equations of the
associated supergravity theories\footnote{In fact the gauge symmetry of IIB Killing spinor equations
is $Spin(9,1)\times U(1)$ but the kappa symmetry projector $\Gamma$ for the D-branes as written in e.g. \cite{eric}
preserves only $Spin(9,1)$ and this suffices for our purpose. Furthermore the IIA supergravity Killing spinor
equations have a hidden $Spin(10,1)$ invariance because they are related to the eleven-dimensional supergravity ones
by dimensional reduction. However
in terms of ten-dimensional variables only the $Spin(9,1)$ gauge group is manifest.}. Therefore gauge symmetry
of the supergravity Killing spinor equations relevant to the spinorial geometry leaves invariant
the associated brane probe supersymmetry conditions (\ref{kappapro}). This can be used to simultaneously solve
 the supergravity Killing spinor equations and the brane supersymmetry conditions.
 We shall use this in the M2-brane  example that we shall present below.

 Next we shall explain the systematics of the supersymmetry condition (\ref{kappapro}), i.e. the way to turn
(\ref{kappapro}) into a linear system which does not contain spacetime gamma matrices
for any spinor $\epsilon$. Let us first consider
the M-branes. The kappa symmetry projector of the M2-brane is
\bea
\Gamma_{M2}={1\over 3!}{1\over \sqrt {|{\rm det}(\gamma)|}} \epsilon^{\mu_1\mu_2\mu_3}
\partial_{\mu_1} X^{M_1} \partial_{\mu_2} X^{M_2}\partial_{\mu_3} X^{M_3}
e^{A_1}_{M_1} e^{A_2}_{M_2}e^{A_3}_{M_3} \Gamma_{A_1 A_2A_3}
\la{kappam2}
\eea
where $\mu_1, \mu_2, \mu_3$ are worldvolume indices, $M_1, M_2, M_3$ are spacetime coordinate indices and $A_1, A_2, A_3$
are spacetime frame indices. In addition,  $\Gamma_A$ are the spacetime gamma matrices, $\Gamma_{A_1 A_2A_3}$ denotes the skew-product of three
gamma matrices, and  $\gamma_{\mu\nu}=\partial_\mu X^M \partial_\nu X^N g_{MN}$ is the induced metric
 on the brane world-volume by pulling back the spacetime metric $g$ with the brane embedding map $X$.
 The kappa symmetry projectors of branes without other worldvolume
and  spacetime fields are similarly
constructed. The above kappa symmetry projector can be written in short-hand notation as
\bea
\Gamma_{M2}=\Phi_{ABC} \Gamma^{ABC}= \Phi_{(3)} \Gamma^{(3)}~.
\eea
Similarly, the M5-brane kappa symmetry projector in the form given in \cite{howe}  can be written as
\bea
\Gamma_{M5}=\Phi_{(2)} \Gamma^{(2)}+\Phi_{(3)} \Gamma^{(3)}+\Phi_{(4)} \Gamma^{(4)}+\Phi_{(5)} \Gamma^{(5)}~,
\eea
where we have used Poincar\' e duality to relate forms with degree $11-k$ to forms of degree $k$, $k<6$, and their
associated Clifford algebra elements. If the M5-brane worldvolume three-form field strength vanishes, then $\Phi_{(2)}=\Phi_{(3)}=\Phi_{(4)}=0$
and so $\Gamma_{M5}=\Phi_{(5)} \Gamma^{(5)}$.

In the context of eleven-dimensional supergravity, one can expand the spinors in a time-like or a null oscillator
basis depending on whether one investigates Killing spinors representing the orbit $SU(5)$ or the orbit
$(Spin(7)\ltimes \bR^8)\times \bR$ of $Spin(10,1)$ in the spinor representation $\Delta_{{\bf 32}}$ \cite{bryant, jose}. The choice between
these two bases
is made purely for convenience as the final result is independent from the basis used. To construct the
linear system associated with the supersymmetry conditions for the M-branes in the time-like basis, as for the supergravity
Killing spinor equations, one separates
the gamma matrix $\Gamma_0$ along the time-direction from the rest. In such case, the kappa symmetry projectors
of the M-branes can be written as
\bea
\Gamma_{Mp}=\sum _k \Psi_{(k)} \tilde\Gamma^{(k)} + \sum _k {\cal X}_{(k)} \tilde\Gamma^{(k)} \Gamma_0
\eea
summing over appropriately chosen $k$, e.g. for the M2-brane we have
\bea
\Gamma_{M2}= \Psi_{(3)} \tilde\Gamma^{(3)}+ {\cal X}_{(2)} \tilde\Gamma^{(2)} \Gamma_0~,
\eea
where $\tilde\Gamma^{(k)}$ denote skew-symmetric k-products of the remaining ten $\Gamma^i$, $i=1,\dots, 10$, gamma matrices.
A $Spin(10,1)$ Majorana Killing spinor can
be written in terms of forms as
\bea
\epsilon&=& f (1+e_{12345})+ i
g (1- e_{12345}) +\sqrt{2} u^i (e_i+
\frac{1}{4!}\epsilon_i{}^{jklm} e_{jklm}) + i \sqrt{2} v^i
(e_i-\frac{1}{4!}\epsilon_i{}^{jklm} e_{jklm}) \nonumber \\ &&+
\frac{1}{2} w^{ij} (e_{ij}-\frac{1}{3!} \epsilon_{ij}{}^{klm}
e_{klm}) + \frac{i}{2} z^{ij}(e_{ij}+\frac{1}{3!}
\epsilon_{ij}{}^{klm} e_{klm})~, \la{genspinor}
\eea
where $f,g,
u^i, v^i, w^{ij}$ and $z^{ij}$ are real spacetime functions. We use the spinor conventions of \cite{systema}. Since $\Gamma_{Mp}$ is a linear
operator to compute $\Gamma_{Mp}\epsilon$, it suffices to evaluate $\Gamma_{Mp} e_{i_1\dots i_I}$
on the six
types of spinors $e_{i_1 \cdots i_I}$ with $I=0,\ldots,5$. In particular, we have
\bea
\Gamma_{Mp} e_{i_1\dots i_I}=\sum _k \Psi_{(k)} \tilde\Gamma^{(k)} e_{i_1\dots i_I} + i(-1)^I
 \sum _k {\cal X}_{(k)} \tilde\Gamma^{(k)} e_{i_1\dots i_I}~.
\eea
The expansion of $\tilde\Gamma^{(k)} e_{i_1\dots i_I}$ for $0\leq k <6$ in terms of the holomorphic ``canonical''
basis $\{1, \Gamma^{\bar\a}1,  \dots, \Gamma^{\bar\a_1\bar\a_2\bar\a_3\bar\a_4\bar\a_5}1\}$, $\a_1, \dots, \a_5=1,\dots,5$,
can be read either from the universal formulae
or from the results that have been presented in the appendices of \cite{systema}.

 In the null basis the analysis is similar. In this case, one separates the gamma matrix along the tenth spatial
 direction denoted by $\Gamma_\nat=- \Gamma_0 \Gamma_1 \cdots \Gamma_9$ from the rest, see \cite{systema}. The kappa symmetry
 projection operators of the M-branes
  can be written as
\bea
\Gamma_{Mp}=\sum _k \Psi_{(k)} \tilde\Gamma^{(k)} + \sum _k {\cal X}_{(k)} \tilde\Gamma^{(k)} \Gamma_\nat~,
\eea
where now $\tilde\Gamma^{(k)}$ denotes the shew-symmetric k-products of the remaining ten, $\Gamma^i$, $i=0,\dots, 9$, gamma matrices. Again
it suffices to evaluate $\Gamma_{Mp}$ on the basis $e_{i_1\dots i_I}$ for $0\leq k <6$. In particular, one finds that
\bea
\Gamma_{Mp} e_{i_1\dots i_I}= \sum _k \Psi_{(k)} \tilde\Gamma^{(k)} e_{i_1\dots i_I} +(-1)^{I+1}
\sum _k {\cal X}_{(k)} \tilde\Gamma^{(k)} e_{i_1\dots i_I}~.
\eea
The expansion of $\tilde\Gamma^{(k)} e_{i_1\dots i_I}$ for $0\leq k <6$ in terms of the ``canonical'' null oscillator basis
$\{1, \Gamma^{\bar a_1} 1, \dots, \Gamma^{\bar a_1\dots \bar a_5} 1\}$, $\bar a_1, \dots, \bar a_5= +, 1, \dots, 4$
 can be read  from the universal formulae of \cite{systema}.

 The construction of the linear system associated  with the supersymmetry condition (\ref{kappapro})  for  type II branes can be
 done in a way similar to that for the M-branes we have explained above. The Majorana-Weyl spinors, $\Delta^\pm_{{\bf 16}}$, of $Spin(9,1)$
 can be expanded in a null oscillator basis. The most general Killing spinor of IIB supergravity can be written as
\bea
\epsilon= p 1+ q e_{1234}+ u^i e_{i5}+{1\over2} v^{ij} e_{ij}+{1\over6} w^{ijk} e_{ijk5}~,
\la{spinoriib}
\eea
where $p,q,u,v$ and $w$ are complex functions on the spacetime, and $i,j,k=1,2,3,4$. Our spinor conventions and the construction
of the null basis can be found in \cite{systemb}. The kappa symmetry projectors
of IIB branes, which we denote collectively by $\Gamma_{IIB}$, can be expanded as
\bea
\Gamma_{IIB}= \sum_{k} \Phi_{(k)} \Gamma^{(k)}~,
\eea
where $\Phi_{(k)}$ may also carry internal indices, e.g. $SL(2,\bR)$ indices for type IIB D-branes.
After an analysis similar to that
we have presented for the M-branes, the evaluation of $\Gamma_{IIB}\epsilon$ for $\epsilon$ given in (\ref{spinoriib}) reduces to the evaluation
of $\Gamma^{(k)}\sigma_I$, $0\leq k\leq 5$, where $\sigma_I$ are the five types of spinors $1, e_{1234}, e_{i5}, e_{ij}$ and $e_{ijk5}$.
Again, the expressions for $\Gamma^{(k)}\sigma_I$ in terms of the
``canonical'' null oscillator basis
$\{1, \Gamma^{\bar a_1} 1, \dots, \Gamma^{\bar a_1\dots \bar a_5} 1\}$, $\bar a_1, \dots, \bar a_5= +, 1, \dots, 4$
can be either be computed from the universal formulae or can be read
from the appendices of \cite{systemb}.

Next let us turn to supersymmetry conditions for  type IIA branes. The Killing spinors of IIA supergravity can be written as
\bea
\epsilon&=&f 1+ g e_{1234}+ u^i e_{i5}+{1\over2} v^{ij} e_{ij}+{1\over6} w^{ijk} e_{ijk5}
\cr
&&~~~~~~~~~~~+\tilde f e_5+ \tilde g e_{12345}+
\tilde u^i e_{i}+{1\over2} \tilde v^{ij} e_{ij5}+{1\over6} \tilde w^{ijk} e_{ijk}~,
\la{spinoriia}
\eea
where the components $f,g, u,v,w$ and $\tilde f, \tilde g, \tilde u, \tilde v,\tilde w$ are real spacetime functions.
The kappa symmetry projectors
of type IIA branes, which we denote collectively by $\Gamma_{IIA}$, can be expanded as
\bea
\Gamma_{IIA}=\sum_k\Phi_{(k)} \Gamma^{(k)}~.
\eea
To evaluate $\Gamma_{IIA}\epsilon$, where $\epsilon$ is given in (\ref{spinoriia}), suffices
to evaluate $\Gamma^{(k)}\sigma_I$ and $\Gamma^{(k)}\Gamma_5\sigma_I$,
where $\sigma_I$ denotes the five types of spinors as in the IIB case above. The expression for $\Gamma^{(k)}\sigma_I$
can be computed as in the IIB case. The expressions for $\Gamma^{(k)}\Gamma_5\sigma_I$ can be found from
those of $\Gamma^{(k)}\sigma_I$ after exchanging the light-cone directions $-\leftrightarrow +$.

To illustrate the use of the spinorial geometry method to solve (\ref{kappapro}),
consider a M2-brane probe  propagating in a supersymmetric supergravity  background preserving at least one supersymmetry.
Other examples will be presented elsewhere \cite{peter}.
In addition, let us assume that the Killing spinor that  satisfies  (\ref{kappapro})
 has stability subgroup $SU(5)$ in $Spin(10,1)$.
Since  as we have explained both the Killing spinor equations of eleven-dimensional supergravity
 and the supersymmetry condition (\ref{kappapro}) are invariant
under $Spin(10,1)$,  one can choose $\epsilon$ to be any representative of the $SU(5)$ orbit in $\Delta_{{\bf 32}}$.
In particular, one can choose $\epsilon$ as in \cite{jug}, i.e.
\bea
\epsilon= f (1+e_{12345})~,
\eea
where $f$ is a spacetime function.
In turn,  the supersymmetry condition (\ref{kappapro}) reads
\bea
\Gamma_{M2}(1+e_{12345})= 1+e_{12345}~.
\la{tmem}
\eea
Since the Killing spinor represents the $SU(5)$ orbit of $Spin(10,1)$ in $\Delta_{{\bf 32}}$, it is convenient to
use the time-like basis to do the analysis.
Separating the gamma matrix $\Gamma_0$ from the rest and expanding (\ref{tmem})  in the canonical holomorphic spinor basis
$\{1, \Gamma^{\bar\a}1,  \dots, \Gamma^{\bar\a_1\bar\a_2\bar\a_3\bar\a_4\bar\a_5}1\}$, $\a_1, \dots, \a_5=1,\dots,5$, we find that
\bea
-2i{\cal X}^{\alpha}{}_{\alpha}&=& 1\nonumber \\
\Psi^{\bar\beta\alpha}{}_{\alpha} &=& 0~,
\nonumber \\
\Psi^{\g_1\g_2\g_3}\,  \tilde\epsilon_{\g_1\g_2\g_3}{}^{\bar\a\bar\b} + 2i{\cal X}^{\bar\a\bar\b} &=&0~,
\label{m2su5}
\eea
where $\tilde \epsilon=\sqrt {2}\, \epsilon$ and $\epsilon$ is the holomorphic Levi-Civita tensor.
In the analysis so far we have not used that $\epsilon$ is a Killing spinor.
To continue, the  spacetime metric of the eleven-dimensional background with Killing spinor $\epsilon= f (1+e_{12345})$
can be written as
\bea
ds^2= - f^4 (dt+\alpha)^2+ ds^2_{10}
\eea
where ${\partial/\partial t}$ is a Killing vector field. The spacetime is a fibre bundle with fibre the orbits of ${\partial/\partial t}$ and
 base space an almost
Hermitian manifold $B$ with metric $ds^2_{10}$ and compatible $SU(5)$ structure. In particular $B$ admits
a Hermitian form $\omega$. The analysis of the geometry of eleven-dimensional backgrounds with one supersymmetry
has originally been done in \cite{pakis} using G-structures but here we follow the spinorial analysis of \cite{jug}.
Next introduce  the spacetime frame $e^0= f^2 (dt+\alpha)$ and $e^{\bar \a}, e^\a$, $\a=1, \dots, 10$, and write
 $ds^2_{10}=
2\delta_{\a\bar\b} e^\a e^{\bar \b}$ and $\omega=-i \delta_{\a\bar\b}\, e^\a\wedge e^{\bar \b}$.
Assuming that the M2-brane world-volume has topology $C\times \bR$, where $C$ is a two-dimensional
surface, and  that ${\partial/\partial t}$ is a rotation free Killing vector field, $d\alpha=0$, which also leaves
invariant the supergravity 4-form field strength, one can set  $X^0=t$,
$\partial_\tau X^\a=0$; $\tau$ is the worldvolume time coordinate.  So we choose a static embedding such that
$C\subset B$. Then, one can see that the only non-trivial conditions are
\bea
-2i{\cal X}^{\alpha}{}_{\alpha}&=& 1~,
\cr
{\cal X}^{\bar\a\bar\b} &=&0~.
\eea
The first equation can be rewritten as
\bea
\omega|_C=d{\rm vol}(C)~,
\eea
where $\omega|_C$ is the hermitian form of $B$ restricted on $C$ and $d{\rm vol}(C)$ is the induced volume on $C$. This is precisely the
condition expected from an almost hermitian generalized calibration \cite{GPT, GIP}, i.e. that the restriction of the hermitian
form on the cycle is equal to its volume. The second condition can be written
as
\bea
d X^{\bar\a}\wedge  d X^{\bar\b}=0~,
\eea
which is clearly satisfied iff the embedding map $X$ is pseudo-holomorphic. Therefore the supersymmetry condition
(\ref{kappapro})  specifies both the calibration type of the supersymmetric submanifolds of the spacetime and
 the  embedding map $X$ of the brane world-volume into the spacetime.

As we have seen in the example above, the equations that arise from the supersymmetry condition (\ref{kappapro}) are generalized
calibrations. Initially the generalized calibrations where proposed as the supersymmetric cycles for branes  embedded in backgrounds
with non-vanishing gauge potentials that induce Wess-Zumino type of couplings in the world-volume brane actions \cite{GPT}.
In particular, world-volume fields, like the Born-Infeld field for D-branes, were required to vanish.
Using  generalized
calibrations, one can establish a bound for the  the energy of brane solitons. Extensions
of the definition of the generalized calibrations  have also been proposed to include
other non-vanishing world-volume fields, like Born-Infeld
type of fields,
 in some special cases, see e.g. \cite{lambert, koerber, smyth}. However, it is clear that if one is interested in all the
 supersymmetric brane configurations, these are given by
solving (\ref{kappapro}). So in the context of branes, one can define as generalized calibrated cycles the solutions
of (\ref{kappapro}). In addition, as we have demonstrated using the spinorial geometry techniques, one can derive the
conditions that these generalized calibrated  cycles satisfy in all cases and without any additional restriction
on either the world-volume or spacetime fields apart from those required by spacetime supersymmetry. One can also formulate a bound
for the Euclidean action of branes in direct analogy to that of generalized calibrations mentioned above.
We shall demonstrate this for Euclidean D-branes, see e.g. \cite{strominger, joseb, marino}, and it easily
 extended to other cases. To see this, suppose that the kappa
symmetry projector is hermitian $\Gamma^\dagger=\Gamma$ and $\Gamma^2=1$ and  that the supersymmetry condition
is given as in (\ref{kappapro}). We also write the (Euclidean) D-brane action \cite{daa} as
\bea
S_{Dp}=\int_W d^{p+1}\sigma\, \sqrt{|{\rm det}(\gamma+{\cal F})|}+S_{WZ}~,
\eea
where ${\cal F}=F-B$ and  $S_{WZ}=\int_W C\wedge e^{\cal F}$ is the Wess-Zumino term. The kappa symmetry projector of \cite{eric} for D-branes
can be written as
\bea
\Gamma={1\over  \sqrt{|{\rm det}(\gamma+{\cal F})|}}\, \hat \Gamma~,
\eea
where $\hat \Gamma$ is a Clifford algebra element that depends of the embedding maps, the Born Infeld field $F$ and the
pull back of $B$. Using a supersymmetry argument, define the spinor $\eta={1\over \sqrt{2} }(1-\Gamma)\epsilon$ and observe that
\bea
0\leq \eta^\dagger \eta= \epsilon^\dagger (1-\Gamma)\epsilon~.
\eea
Multiplying the above expression with  $\sqrt{|{\rm det}(\gamma+{\cal F})|}$, assuming that it does not vanish, we find that
 \bea
\sqrt{|{\rm det}(\gamma+{\cal F})|}\geq {1\over  \epsilon^\dagger \epsilon}\, \epsilon^\dagger \hat \Gamma \epsilon~.
\eea
Integrating the above expression over the brane world-volume and adding  $S_{WZ}$, we find\footnote{We have not normalized $\epsilon$,
as $\epsilon^\dagger \epsilon=1$, because
it is not apparent in the context of supergravity with fluxes that that the normalized spinor of a Killing spinor is Killing.}   that
\bea
S_{Dp}\geq \int_W \phi + S_{WZ}~,~~~\phi={1\over  \epsilon^\dagger \epsilon}\,\, \epsilon^\dagger \hat \Gamma \epsilon~.
\la{boundp}
\eea
 Observe that $\phi$ is a (p+1)-form
which depends on the spacetime Killing spinor bi-linears as well as the brane world-volume $F, X$ and spacetime fields $g$ and $B$.
A similar bound for M5-brane solitons has been given in \cite{lambert}.
In the absence of worldvolume and spacetime fluxes, the bound (\ref{boundp}) reduces to that of standard calibrations
\cite{harvey}, see also \cite{strominger}.
Therefore it is tempting to identify $\phi$ with the calibration form of this type of generalized calibrations.
  In the context of generalized calibrations of \cite{GPT}, the generalized calibration form is related to the supergravity gauge potentials.
  If this applies here, then $\phi$ must be equal to the combination of the supergravity gauge potentials that appear in $S_{WZ}$
  of the Dp-brane action.
This bound is attained, iff $\eta=0$ and so (\ref{kappapro}) is satisfied.

\vskip 0.5cm
{\bf Acknowledgments:}~G.P. thanks Ulf Gran, Jan Gutowski, and Diederic Roest for many helpful discussions on the spinorial geometry
and C. Bachas for comments on calibrations.

\end{document}